\documentclass{emulateapj}

\input{epsf}

\submitted{Accepted by ApJ Letters on May 8, 2008}

\shorttitle{NGC 5128 as an isotropic rotator}

\shortauthors{Ewa L. \L okas}

\begin{document}

\title{NGC 5128 as an isotropic rotator}

\author{Ewa L. \L okas}

\affil{Nicolaus Copernicus Astronomical Center, Bartycka 18, 00-716 Warsaw, Poland; lokas@camk.edu.pl}


\begin{abstract}
NGC 5128 is a well-studied elliptical galaxy with an excellent kinematic data set
for planetary nebulae which has been modelled up till now only by solving the Jeans
equation for spherical systems. As a first approximation beyond spherical symmetry
we model the galaxy as an axisymmetric system flattened by rotation
with isotropic velocity distribution.
We propose a new version of such an isotropic rotator having cuspy density profile
with NFW-like behavior. The solutions of the Jeans equations for
a single component of such form do not reproduce the data
well: the rotation curve rises too slowly with radius and the velocity dispersion
profile drops too fast. The data are well fitted however by a system built with two
components: a more compact, flattened, less massive, fast-rotating and cold `bulge' and
an extended, almost spherical, more massive, slow-rotating and hot `halo'. This picture agrees well
with the results of recent N-body simulations of galaxy mergers which tend to
produce oblate rotating spheroids. The total mass of the
system is estimated to be $(9.1 \pm 3.5) \times 10^{11}$ solar masses
and the mass-to-light ratio is only $26 \pm 17$ solar units.
\end{abstract}



\section{Introduction}

NGC 5128 (Centaurus A) is a well-studied nearby elliptical galaxy (for a review see
Israel 1998) with an excellent kinematic data set of 780 planetary nebulae (PNe) with
measured velocities (Peng, Ford, \& Freeman 2004). The data extend out to about
16 effective radii allowing us to probe the dynamics of the galaxy not only in the
usual inner part but well beyond the distribution of light. Peng et al. found that the
zero velocity curve deduced from the data departs from the minor axis and concluded that
the galaxy must be triaxial. However, they modelled the PNe kinematics only via the
spherical Jeans equation.

Since the rotation is much stronger along the major axis than the
minor axis it is reasonable to consider, as a first approximation beyond spherical
symmetry, an axisymmetric model in which the galaxy is an oblate spheroid
flattened by rotation. This conjecture is strongly supported by the results of recent
simulations of dissipational mergers of disks where oblate systems are the most frequent
outcomes (Cox et al. 2006). In addition, the observed rotation along the minor axis
could be at least in part due to contamination from a merger remnant (now visible as
the spectacular dust lane and gas ring along the minor axis of the galaxy) with its own population
of PNe not in equilibrium
with the rest of the galaxy (Sparke 1996). The measured kinematics of both neutral
(Schiminovich et al. 1994) and ionized (Bland, Taylor, \& Atherton 1987) hydrogen in this region
shows that the NW side of the ring is approaching while the
SE one is receding, like the PNe along the minor axis.

A number of analytic potential-density pairs for axisymmetric systems can be found
in the literature. Miyamoto \& Nagai (1975) and Satoh (1980) proposed a family of generalized
Plummer models that can describe axisymmetric galactic systems with a combination
of random motions and rotation. The disadvantage of these models is that their density distributions
possess cores and therefore are unlikely to reproduce well the density profiles of elliptical
galaxies (Mamon \& {\L}okas 2005a,b). Other models that could be generated from the well-known
potentials by rescaling the cylindrical $z$-coordinate tend to produce non-physical negative
densities along the rotation axis (Binney \& Tremaine 1987).

In this work we propose a new potential generating a cuspy axisymmetric density distribution with
an NFW-like (Navarro, Frenk, \& White 1997) asymptotic behavior. We use this model in solving the
Jeans equations assuming isotropic velocity distribution and finding the projected rotation velocity
and dispersion profiles. The solutions are then fitted to the data for PNe in NGC 5128. The new model
allows us to consider for the first time different inclinations and flattenings and constrain the
mass of the galaxy without the assumption of spherical symmetry.


\section{A new isotropic rotator}

We propose to approximate the (positive) axisymmetric potential of the galaxy with a formula
\begin{equation}                         \label{potential}
	\Phi (R, z) = \frac{G M}{b + (1/2) f_1 + (1/2) f_2}
\end{equation}
with
\begin{eqnarray}
	f_1 (R, z) &=&  (R^2 + z^2)^{1/2}
	\nonumber  \\
	f_2 (R, z, a, b) &=&
	\{ R^2 + [a + (b^2 + z^2)^{1/2}]^2 \}^{1/2}
	\nonumber
\end{eqnarray}
where $R$ and $z$ are the standard cylindrical coordinates, $M$ is the total mass of the system,
while $a$ and $b$ are constants with the dimension of length. The form of the potential
was motivated on one hand by the Hernquist (1990) model and on the other by the way
suggested by Miyamoto \& Nagai (1975) and Satoh (1980) to flatten Plummer potentials.
The formula for the density distribution following
from the potential (\ref{potential}) via the Poisson equation is rather complicated but has
NFW-like asymptotic behavior along $R$, with $R^{-1}$ at low $R$ and $R^{-3}$ slope at large $R$. Along the
$z$-axis the density decreases more steeply, like $z^{-1}$ at low $z$ and $z^{-4}$ at large $z$.

As in Miyamoto \& Nagai (1975) and Satoh (1980) models, the ratio $b/a$ controls the shape
of the system and the fraction of kinetic energy
associated with the random versus ordered motion, so that systems with $b/a \ll 1$ are flattened,
fast rotators with little random motion (spiral galaxies), while those with $b/a \gg 1$ are almost
spherical systems dominated by random motions with little rotation (elliptical galaxies). An interesting
intermediate case is when $b/a$ is of the order of a few; such systems are only moderately
flattened and possess similar amount of random and ordered motion.

We denote the velocities along the three cylindrical coordinates $R$, $\theta$ and $z$ respectively
by $u=v_R$, $v = v_\theta$ and $w = v_z$. Their variances are related by the set of Jeans
equations (Binney \& Tremaine 1987)
\begin{eqnarray}
	\frac{\partial}{\partial R} (\rho \langle u^2 \rangle )
	+ \frac{1}{R} (\rho \langle u^2\rangle  - \rho \langle v^2 \rangle ) &=&
	\rho \frac{\partial}{\partial R} \Phi
	\label{jeans1}  \\
	\frac{\partial}{\partial z} ( \rho \langle w^2 \rangle ) &=&
	\rho \frac{\partial}{\partial z} \Phi.
	\label{jeans2}
\end{eqnarray}
We assume that the only streaming motion in the galaxy is the rotation so $ \langle u \rangle =
\langle w \rangle =0$, $\langle v \rangle \neq 0$ and the residual velocity components around the
mean are isotropic everywhere in the system: $ \langle u^2 \rangle = \langle w^2 \rangle =
\langle v^2 \rangle - \langle v \rangle^2$.

For comparison with observations the quantities discussed above have to be projected along the
line of sight and on the plane of the sky. We express the projected quantities in terms of
Cartesian coordinates $X, Y$ and $L$ centered on the models with $X$-axis along the projected major
axis of the galaxy, $Y$ along the minor axis, and $L$ along the line of sight. The inclination
angle $i$ is the angle between the equatorial plane of the galaxy and the line of sight.
The projected surface density distribution, mean rotational velocity and velocity dispersion
are then given respectively by
\begin{eqnarray}
	\mu (X,Y) &=& \int_{-\infty}^{+\infty} \rho \; {\rm d} L \label{projected1}  \\
	V (X,Y) &=& \frac{1}{\mu} \int_{-\infty}^{+\infty} \nu \rho \langle v \rangle
	{\rm d} L \label{projected2}  \\
	\sigma^2 (X,Y) &=& \frac{1}{\mu} \int_{-\infty}^{+\infty} \rho \langle w^2 \rangle {\rm d} L
	\label{projected3}
\end{eqnarray}
where $\nu=X \cos i/[X^2+(Y \sin i -L \cos i)^2]^{1/2}$ is the direction cosine.

\section{The data}

The data comprise 780 spectroscopically confirmed PNe from Peng et al. (2004, including the earlier data
set of Hui et al. 1995) with positions and velocities measured with high accuracy so that the velocity
errors can be neglected in the further analysis. Plots of velocities versus positions do not show any
outliers which suggests that the data set is not contaminated by any background or neighboring systems
and all PNe can be considered as members of the galaxy.

\begin{figure}
\begin{center}
    \leavevmode
    \epsfxsize=8.5cm
    \epsfbox[105 110 330 405]{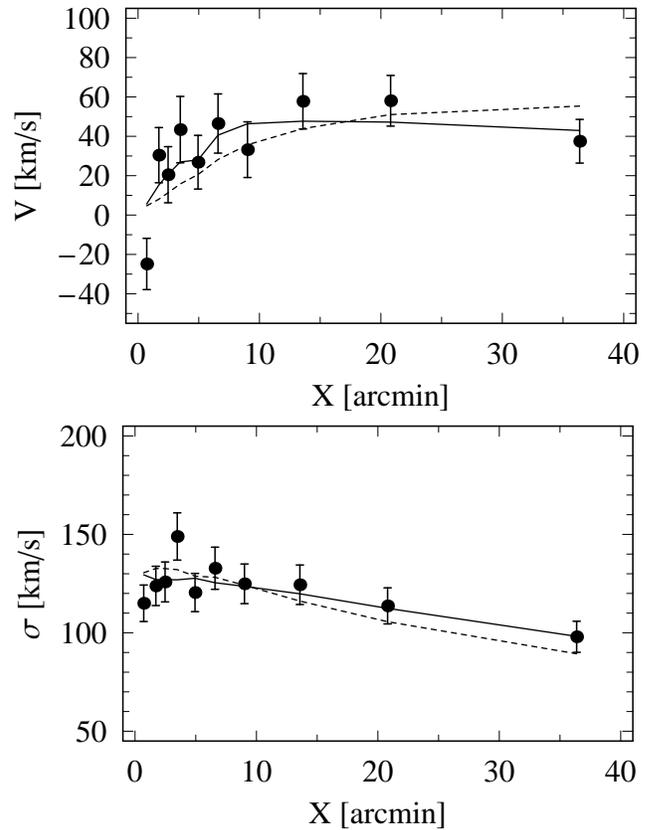}
\end{center}
\figcaption{The mean projected rotation velocity (upper panel) and the line-of-sight velocity
dispersion profile (lower panel) of PNe in NGC 5128. In both panels the dashed lines show the
best-fitting solutions of the Jeans equations for isotropic one-component model and the solid lines
for the two-component model. \label{moments}}
\end{figure}


The data were transformed using the symmetries of the axisymmetric system so that all
PNe have positive $X$, $Y$ positions. It was assumed that the minor axis of the optical image of the
galaxy is the projected rotation axis and the major axis has $PA=35^\circ$ (Dufour et al. 1979).
The PNe were then ordered by their position along the major axis
($X$) and binned with $n=78$ objects per bin. This resulted in 10 data points for both line-of-sight
rotation velocity and dispersion at mean $(X, Y)$ values in the bin. The dispersions
$\sigma$ were estimated using the standard unbiased estimator and assigned
sampling errors of size $\sigma/\sqrt{2(n-1)}$. The mean velocities were assigned
standard errors of the mean $\sigma/\sqrt{n}$. The velocity moments estimated from the
data are plotted in Figure~\ref{moments} as a function of $X$. The moments are measured close
to the major axis since mean $Y < 8'$ for all ten data points.

The velocity dispersion profile presented in Figure~\ref{moments} shows remarkable similarity to
those of regular clusters of galaxies (e.g. {\L}okas \& Mamon 2003;
{\L}okas et al. 2006; Wojtak \& {\L}okas 2007) and theoretical predictions for objects with
NFW-like density distributions and isotropic orbits ({\L}okas \& Mamon 2001).
The situation with rotation velocity is less clear: the innermost point
falls well below zero velocity expected in the center, which we suspect to be due to the
contamination from the dust lane-related merger remnant.



\section{The models}

The data shown in Figure~\ref{moments} were fitted with the projected solutions of the Jeans equations
(\ref{projected2})-(\ref{projected3}) assuming that the potential in the galaxy is given by formula
(\ref{potential}) and the density follows from it via Poisson equation.
Our approach here is rather unorthodox compared e.g. to
Binney, Davies, \& Illingworth (1990) where the distribution of light was used to constrain potential.
However, since population gradients are present in the galaxy it is not obvious that mass follows light
in this component. In addition, the distribution of light in NGC 5128 is difficult to estimate along
the minor axis where it is obscured by the prominent dust lane. We therefore proceed as if no information
was available on the stellar mass. In principle, the density $\rho$ in equations
(\ref{jeans1})-(\ref{projected3}) refers to the tracer (PNe) whose kinematics we study. We do not estimate it
from the distribution of PNe since their number is low and sampling not uniform, it does not have to follow the
light either. In any case, the distribution of the tracer is not critical to the modelling because it partially
cancels out due to the way it enters equations (\ref{projected1})-(\ref{projected3}). For simplicity we
suppose here that PNe follow the total density.

Both rotation velocity and dispersion data can be fitted simultaneously
by simple $\chi^2$ minimization since
they are very weakly correlated (as can be shown by Monte Carlo sampling from nearly Gaussian
distributions analogous to the one described by {\L}okas \& Mamon 2003). The adopted
distance to the galaxy was $D=3.84 \pm 0.35$ Mpc (Rejkuba 2004) so that $1'$
corresponds to 1.12 kpc.


\begin{table}
\begin{center}
\caption{Fitted parameters of the two-component models \label{double}}
\begin{tabular}{cccccccc}
\tableline\tableline
$i$   & $a$      & $b$      & $M_1$                 &   $c$    &    $d$    & $M_2$   & $\chi^2/N$ \\
\tableline
0  & 2.2  & 2.7 & 1.1 & 2.2 & 15.7 & 8.2 & 19/13 \\
10 & 2.5  & 4.0 & 1.4 & 1.2 & 15.7 & 7.8 & 18/13 \\
20 & 2.8  & 4.1 & 1.5 & 1.6 & 15.5 & 7.6 & 18/13 \\
30 & 3.9  & 4.3 & 1.5 & 1.1 & 15.2 & 7.4 & 17/13 \\
40 & 5.3  & 4.1 & 1.6 & 1.1 & 14.9 & 7.3 & 17/13 \\
50 & 8.0  & 3.4 & 1.7 & 1.6 & 14.3 & 7.4 & 17/13 \\
60 & 9.2  & 2.8 & 0.3 & 2.7 & 11.1 & 9.1 & 18/13 \\
70 & 9.2  & 1.9 & 0.0 & 10.2&  7.6 & 10.1& 21/13 \\
\tableline
\end{tabular}
\tablecomments{The inclination $i$ is in degrees, the scalelengths $a$, $b$, $c$ and $d$ in arcmin and the
masses $M_1$ and $M_2$ in $10^{11}$ M$_\odot$.}
\end{center}
\end{table}

\begin{figure}
\begin{center}
    \leavevmode
    \epsfxsize=8.1cm
    \epsfbox[100 30 380 305]{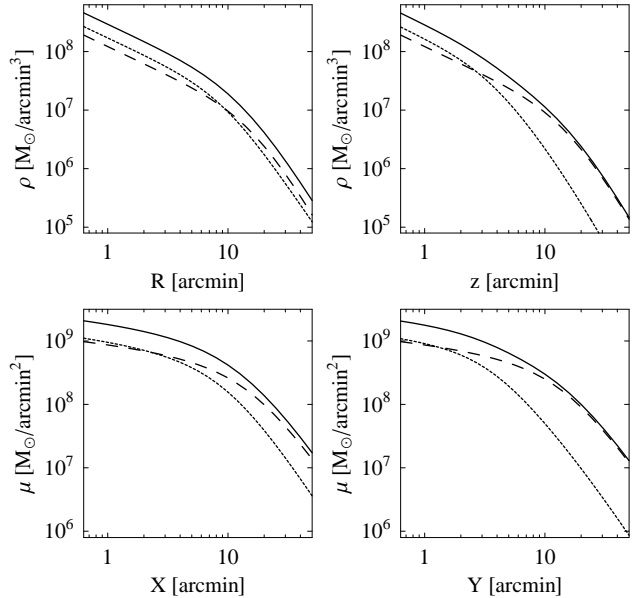}
\end{center}
\figcaption{Density profiles of the best-fitting two-component model for $i=50^\circ$
(see Table~\ref{double} for the parameters). Two upper panels show the 3D density profiles:
in the equatorial plane, $z=0$ (left panel) and along the rotation axis, $R=0$ (right panel).
Two lower panels plot the surface density distributions: along the major axis (left panel) and
along the minor axis (right panel). In each plot
the short-dashed line is for the compact component, the dashed line for the extended
component and the solid
line shows the total density. \label{density}}
\end{figure}

In the case of single-component models the best fit with $\chi^2/N=25.9/16$ is found
for $i=30^\circ$ (with $i=0^\circ$ corresponding to the edge-on view), but similar quality
fits are found for a wide range of inclinations, only the ratio $b/a$ of
the best-fitting parameters decreases systematically with inclination, since the galaxy then has to have
more rotation to make up for the projection effects.

The best-fitting solutions for $V$ and $\sigma$ corresponding to the $i=30^\circ$
case with $a=4.4'$, $b=10.0'$ and $M=7.3 \times 10^{11}$ M$_\odot$
are plotted in Figure~\ref{moments} as dashed lines. The quality of the fits is not satisfactory,
the best solution for the rotation velocity rises too slowly with radius and
the dispersion profile drops too fast. In addition, when the data for rotation velocity and dispersion
are fitted separately good fits are obtained, but for rather different parameter values. While the
rotation curve is well reproduced by a low-mass model with $b/a \approx 1$, the dispersion profile needs
higher mass and $b/a \gg 1$. It is therefore reasonable to expect that both data sets would be better
reproduced by two-component models.



In the case of two-component models
the first component is parametrized by the scalelengths $a, b$ and the total mass $M=M_1$, the second by
analogous quantities $c, d$ and $M_2$. The parameters of the best-fitting models for different
inclinations are listed in Table~\ref{double}. The quality of the fits is now acceptable
($\chi^2/N>1$ only because of the discrepant inner-most velocity data point). Again the $b/a$ ratio
decreases with growing inclination and similar quality fits are obtained for a wide range of
inclinations.

\begin{figure}
\begin{center}
    \leavevmode
    \epsfxsize=8.4cm
    \epsfbox[100 15 380 150]{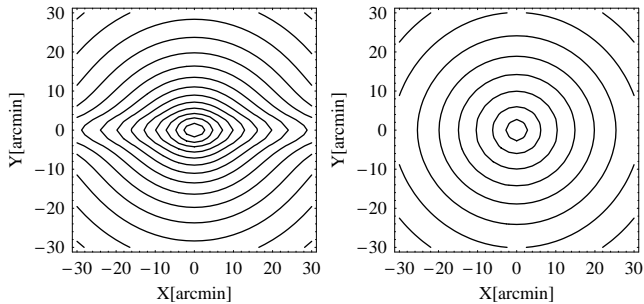}
\end{center}
\figcaption{Surface density distribution $\mu$ of the
compact, flattened component (left panel)
and the extended, almost spherical component (right panel)
for the best-fitting model with $i=50^\circ$ (see Table~\ref{double}).
In both panels the innermost contours are for $\log \mu=8.8$
and the other ones are equally spaced with a step of 0.2 in  $\log \mu$
(with $\mu$ in units of M$_\odot$/arcmin$^2$). \label{components}}
\end{figure}

The best fit with $\chi^2/N=17.3/13$ is found for $i = 50^\circ$. The
solutions for this case are plotted in Figure~\ref{moments} as solid lines.
The density profiles
of the two components and the total density distribution are compared in Figure~\ref{density}.
The surface distribution $\mu(X,Y)$ of both components is presented in Figure~\ref{components}.
The picture emerging from these results is that the velocity distribution of PNe in NGC 5128 can be
well reproduced by a more compact, less massive and faster rotating flattened component
with $b/a \approx 1$ together with a more extended, more massive and slowly rotating almost spherical
component with $d/c \gg 1$. The compact component contributes mostly to the inner part of the rotation
curve while its contribution to the dispersion is small. On the other hand, the extended component
contributes to the rotation mostly in the outer parts but dominates in the dispersion.
The two components can be interpreted as the traditional stellar bulge and dark matter halo
in elliptical galaxies. The comparison between the surface density distribution of the more compact
component and the light distribution in NGC 5128 (Dufour et al. 1979) shows that the latter
(well approximated by the de Vaucouleurs profile with an effective radius of $5'$) is somewhat
less extended, especially along the major axis. Note that the results do not change significantly if
we assume that only one (the more compact) instead of both components contribute to the tracer.

The best-fitting total mass of both components is $M=M_1 + M_2 = (9.1 \pm 3.5) \times
10^{11}$ M$_\odot$ where the 1$\sigma$ error (dominated by the sampling errors
of velocity moments) was estimated from the $\Delta \chi^2$
statistics by Monte Carlo sampling of the parameter
space and marginalizing over the remaining parameters. Adopting the luminosity of the galaxy
(de Vaucouleurs et al. 1991)
$L_V = (3.45 \pm 0.88) \times 10^{10}$ L$_{\odot, V}$ (where the 1$\sigma$ error includes the
error in the estimate of the total apparent brightness as well as the distance)
we get the mass-to-light ratio $M/L_V = 26.3 \pm 16.9$ M$_\odot$/L$_\odot$. The mass-to-light
ratio for the more compact component is a few times
lower, $M_1/L_V = 4.9$ M$_\odot$/L$_\odot$, consistent with stellar values, which confirms that
this component indeed corresponds to the light distribution in the galaxy.

Our mass and $M/L$ estimates are somewhat
larger than those found by Peng et al. (2004) by solving the Jeans
equation for spherical systems. Within 80 kpc Peng et al.
estimate the total mass to be 5-6 $\times 10^{11}$ M$_\odot$ while at the same distance we find
$M=6.6 \times 10^{11}$ M$_\odot$. The values agree within our mass error which was found
to be larger than 30 percent.

\section{Conclusions}

The total mass-to-light ratio estimated here is rather low,
in concordance with the results claiming the dearth of dark matter by Romanowsky et al. (2003), but
in disagreement with the high $M/L$ values found from the analysis of dispersion profiles
of satellites of isolated elliptical galaxies by Prada et al. (2003) and the results of
N-body simulations (e.g. Marinoni \& Hudson 2002).
As discussed by Mamon \& {\L}okas (2005b) and Dekel et al. (2005) the disagreement can be weakened by
considering radially anisotropic, tidally affected stellar orbits and different density profiles.
Note however, that the results of Romanowsky et al. as well as and the proposed remedies were
based on or trying to explain the velocity dispersion profiles of PNe strongly decreasing with radius,
while in NGC 5128 the dispersion decreases in the outermost bin by no more than 30 percent with
respect to the central value.

The new analysis of the PNe data for NGC 5128 in the framework of axisymmetric systems presented here
points to the existence of a flattened, rotating component whose mass distribution
follows the light distribution in the galaxy except for the equatorial plane where it is
more extended. The need for the presence of such a
disky component in elliptical galaxies to explain their line-of-sight velocity distribution
has been postulated by Naab \& Burkert (2001) based on N-body simulations of mergers.
The model of NGC 5128 proposed here provides further evidence for the extended rotating component
in elliptical galaxies.


\acknowledgments

This research was supported in part by the Polish Ministry of Science and Higher Education
under grant N N203 0253 33.

\end{document}